\documentclass{PoS}

  \ifx\pdfoutput\undefined
   \usepackage[dvips]{graphicx}
   \else
   \usepackage[pdftex]{graphicx}
   \fi
   \usepackage{epstopdf}

\title{Exclusive meson leptoproduction and spin dependent generalized parton distributions}

\ShortTitle{Meson Production GPDs}

\author{\speaker{Gary R. Goldstein}\thanks{This work is partially support by U.S. Department of Energy Grant DE-FG02-92ER40702.}\\
        Department of Physics and Astronomy,\\
        Tufts University \\
        Medford, MA 02155 USA
        E-mail: \email{gary.goldstein@tufts.edu}}

\author{Simonetta Liuti\\
        Department of Physics, \\
        University of Virginia \\
        Charlottesville, VA 22901 USA \\
        E-mail: \email{sl4y@virginia.edu}}

\abstract{.Exclusive meson leptoproduction from nucleons in the deeply virtual exchanged boson limit can be described by generalized parton distributions (GPDs). Including spin dependence in the description requires 8 independent quark-parton and gluon-parton functions. The chiral even subset of 4 quark-nucleon GPDs are related to nucleon form factors and to parton distribution functions. The chiral odd set of 4 quark-nucleon GPDs are related to transversity, the tensor charge, and other quantities related to transversity. Different meson or photon production processes access different combinations of GPDs. This is analyzed in terms of t-channel exchange quantum numbers, $J^{PC}$ and it is shown that pseudoscalar production can isolate chiral odd GPDs. There is a sensitive dependence in various cross sections and asymmetries on the tensor charge of the nucleon and other transversity parameters.}

\FullConference{XVIII International Workshop on Deep-Inelastic Scattering and Related Subjects, DIS 2010\\
		April 19-23, 2010\\
		Firenze, Italy}

\begin{document}

\section{Introduction - Spin Dependent GPDs}

Deeply virtual exclusive leptoproduction of photons and mesons (DVCS and DVMP) can be described in terms of Generalized Parton Distributions (GPDs). With measurements of polarization and angular asymmetries, a general parameterization requires 8 quark-nucleon spin-dependent GPDs and a corresponding number of gluon-nucleon GPDs (for a review, see ref~\cite{Diehl}). The basic definition of the quark-nucleon GPDs is through off-forward matrix elements of quark field correlators,
\begin{eqnarray}
\label{corr1}
\Phi_{ab} = \int  \, 
\frac{ d y^-}{2 \pi} \,  
e^{i y^- X} \, \langle P^\prime S^\prime \mid \overline{\psi}_b(0)  \psi_a(y^-) \mid P S \rangle
\end{eqnarray}
where we write the Dirac indices explicitly. Contracting with the Dirac matrices, $\gamma^\mu$ or $ \gamma^\mu \gamma^5$ and integrating over the internal quark momenta gives rise to the Chiral even GPDs $H, E$ or $\widetilde{H}, \widetilde{E}$, respectively. On the other hand, contracting with $\sigma^{\mu \nu} $ yields the 4 chiral odd GPDs, $H_T, E_T, \widetilde{H}_T, \widetilde{E}_T$, through
\begin{eqnarray}
\int dk^- \, d^2{\bf k} \,  
 \, \rm{Tr} \left[ i\sigma^{+i} \Phi \right]_{X P^+=k^+} & & \nonumber \\
  =   \frac{1}{2 P^+}  \, \overline{U}(P^\prime, S^\prime) \,
 [ H_T^q \, i \sigma^{+ \, i} + & \widetilde{H}_T^q  \, \frac{P^+ \Delta^i - \Delta^+ P^i}{M^2} & \, 
 +E_T^q \, \frac{\gamma^+ \Delta^i - \Delta^+ \gamma^i}{2M} \,
 \nonumber \\
 &   + \widetilde{E}_T^q \, \frac{\gamma^+P^i - P^+ \gamma^i}{M}]  & U(P,S) 
\label{oddgpd} 
\end{eqnarray}

The crucial connection to the 8 GPDs that enter the partonic  description of electroproduction is through the helicity decomposition~\cite{Diehl}, where, for example, one of the chiral even helicity amplitudes is given by substituting explicit Dirac spinors for nucleons to yield
\begin{equation}
A_{++,++}(X,\xi,t)=\frac{\sqrt{1-\xi^2}}{2}(H^q+{\tilde H}^q-\frac{\xi^2}{1-\xi^2}(E^q+{\tilde E}^q)),
\label{chiraleven}
\end{equation}
\noindent while one of the chiral odd amplitudes is obtained from  Eq.~\ref{oddgpd},
\begin{equation}
A_{++,--}(X,\xi,t)=\sqrt{1-\xi^2}(H_T^q+\frac{t_0-t}{4M^2}{\tilde H}_T^q-\frac{\xi}{1-\xi^2}(\xi E_T^q+{\tilde E}_T^q)).
\label{chiralodd}
\end{equation}

We have constructed a robust model for the GPDs, extending previous work~\cite{AHLT} that is based on diquark spectators and Regge behavior at small $X$. The GPDs are constrained by their relations to PDFs, $H^q(X,0,0)=f_1^q(X)$, ${\tilde H}^q(X,0,0)=g_1^q(X)$, $H_T^q(X,0,0)=h_1^q(X)$ and to nucleon form factors $F_1(t)$,  $F_2(t)$, $ g_A(t)$, $ g_P(t)$ through the first $x$ moments of $H(X,\zeta,t), E(X,\zeta,t), \tilde{H}(X,\zeta,t), \tilde{E}(X,\zeta,t)$, respectively. These are all normalized to the corresponding charge, anomalous moment, axial charge and pseudoscalar ``charge''. For Chiral odd GPDS there are fewer constraints. $H_T(X,0,0)=h_1(X)$ can be fit to the loose constraints in ref.~\cite{Anselm} - the first moment of $H(X,\xi,t)$ is the ``tensor form factor'', called $g_T(t)$ by H\"{a}gler~\cite{Haegler}. Further, it is conjectured that the first moment of $2{\tilde H}_T^q(X,0,0)+E_T^q(X,0,0)$ is a  ``transverse anomalous moment'', $\kappa_T^q$, with the latter defined by Burkardt~\cite{Burk}.

\begin{center}
\begin{figure} 
\begin{center}
\includegraphics[width=.6\textwidth]{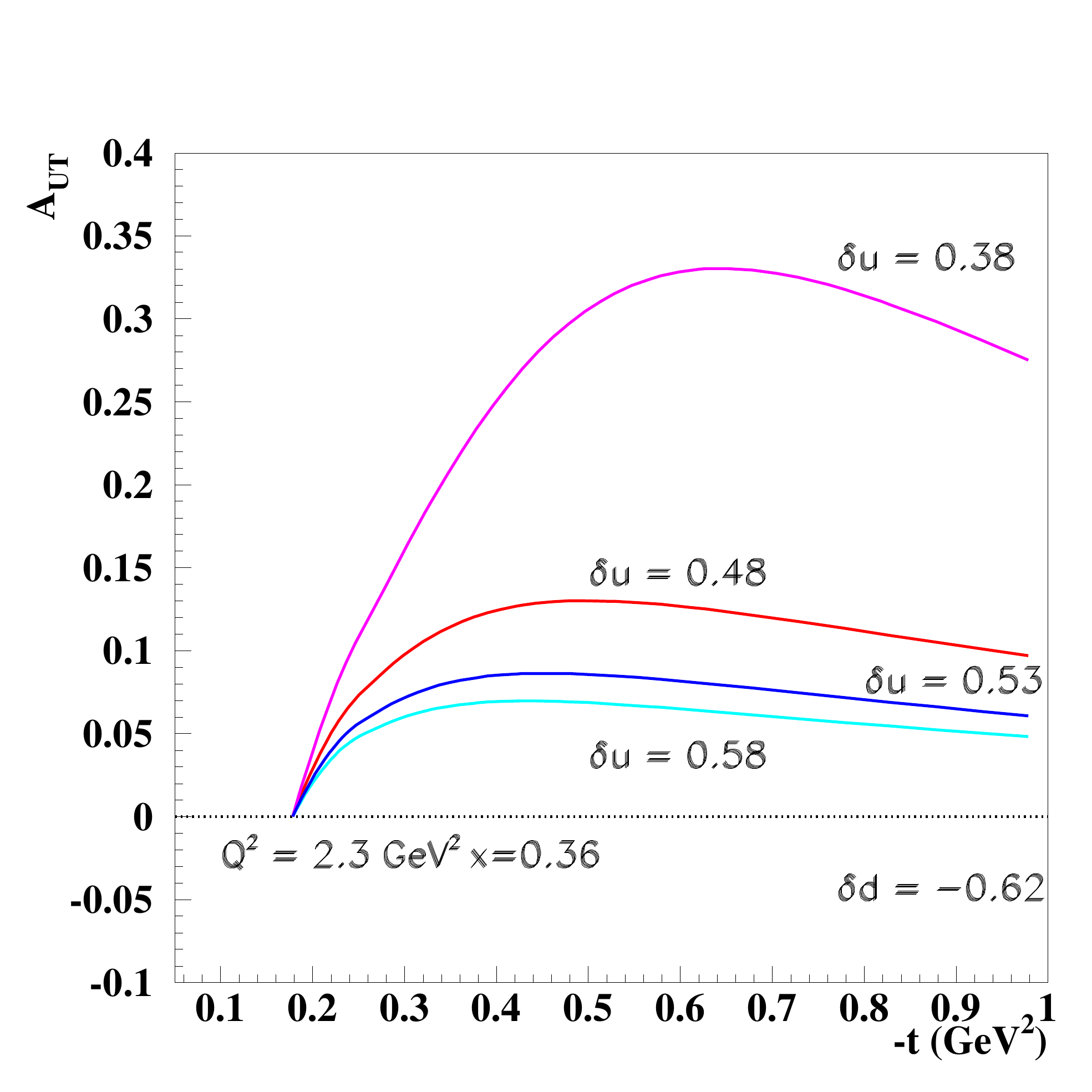} 
\end{center}
\caption{Transverse spin asymmetry, $A_{UT}$,  vs. $-t$, at 
$Q^2=2.3$ GeV$^2$, $x_{Bj}=0.36$ for different
values of the tensor charge, $\delta u$, with fixed 
 $\delta d=-0.62$, {\it i.e.} equal to the central 
value extracted in a global fit~\cite{Anselm}.} 
\label{fig1} 
\end{figure} 
\end{center}

With our {\it ansatz} many observables can be determined in parallel with corresponding Regge predictions. Since the initial work~\cite{Ahmad}, we have undertaken a more extensive parameterization,and  presented several new predictions. Here we show one example -  the transversely polarized target asymmetry, in Fig.~\ref{fig1}.

In the case of $\pi^0$ production there are important constraints that restrict the GPDs that contribute. Consider the t-channel quantum numbers corresponding to combinations of  GPDs. The x moments of the GPDs have expansions in terms of t-dependent form factors and polynomials in $\xi$. It has been shown by Lebed and Ji for pdfs~\cite{LebJi} and Haegler for GPDs~\cite{Haegler}, that these moments have t-channel angular momentum decompositions, as appropriate for t-channel exchanges, as well as Regge poles. 
 From the $t-$channel perspective $\gamma^* + \pi^0$, with C-parity negative, goes into a $q\bar{q}$ pair, which subsequently becomes an $N\bar{N}$ system. In that chain, each part has the same $J^{PC}$. 

Consider first the chiral even GPDs. The crossing odd $\widetilde{H}$ has contributions from $2^{--}, 4^{--}$ and higher. There are several different reasons that this GPD is not expected to contribute at leading order. There cannot be a $0^{--}$ coupling to $\gamma+\pi^0$ or  $N \bar{N}$. The $2^{--}$ appears in the triplet spin with L=2. For simple resonance exchanges, there would be an angular momentum barrier compared to the $J=1$ exchanges. In Regge language the trajectory with $2^{--}$ is non-leading and the absence of $0^{--}$ would require 
a ``nonsense'' factor killing the pole, thereby minimizing the effect in the physical region. The crossing odd $\widetilde{E}$ has contributions from $1^{+-}, 2^{--}, 3^{+-},$ etc., so it is the leading candidate for chiral even GPDs that contribute to $\pi^0$.  Its first moment is the pseudoscalar form factor, for 
which the main contribution is the $\pi$ itself. However, for the neutral $\pi$ this is not the case - there is no $\pi$ pole. How does this effect the $\pi^0$ production?


The electroweak form factors of the nucleon include $g_A(t)$, the axial vector form factor, and $g_P(t)$, the ``induced pseudoscalar" and for  the axial electroweak current, 
\begin{eqnarray}
 \langle N(p^\prime) \mid J^\nu_A \mid N(p) \rangle = {\bar u}(p^\prime)[ g_A(q^2) \gamma^\nu \gamma^5 + \frac{g_P(q^2)}{m_\mu} q^\nu \gamma^5] u(p),
\end{eqnarray} 
where $q^2=(p^\prime - p)^2=t$.
The divergence of the isovector part of the axial current is approximated in PCAC by the pion pole -  the Goldberger-Treiman relation for $g_A(0)$ in terms of the $\pi$-nucleon coupling constant. For non-zero $q^2$ there is a relation between the two form factors,
\begin{eqnarray} 
g_P(q^2)=\frac{2 m_\mu M}{m_\pi^2-q^2} g_A(0).
\label{GTreln}
\end{eqnarray}
Now $\widetilde{H}$ integrates to $g_A(t)$; $\widetilde{E}$ integrates to $h_A(t)=\frac{2 M}{m_\mu} g_P(t)$, proportional to the above pseudoscalar form factor of the nucleon. Recent experimental determinations show that $g_A(0) =1.267$ and $g_P(-0.88m_\mu^2)= 8.58$~\cite{GorFea}. 
For $\pi^0$ electroproduction on the nucleon, however, there is no $\pi^0$ exchange. 
The difference between $g_P$ from Eq.~\ref{GTreln} and experiment is a measure of the non-pole contribution, which is quite small~ \cite{GorFea}.
Thence, the size of  $\pi^0$ electroproduction cross sections would be expected to be considerably less than charged $\pi$'s if $\widetilde{E}$ were the major contribution.

\section{$\pi^0$ and pseudoscalar production}
The measured cross section for $\pi^0$ is sizable and has large transverse $\gamma^*$ contributions. This indicates that the main contributions should come from chiral odd GPDs, for which the t-channel decomposition is richer. In particular, because these GPDs arise from the Dirac matrices $\sigma^{\mu \nu}$, there are 2 series for each GPD corresponding to space-space or time-space combinations - $1^{--}$ and $1^{+-}$. These series occur for 3 of the 4 chiral odd GPDs, the exception being $\widetilde{E}_T$. We are thus led to the conclusion that chiral odd GPDs will dominate the neutral pseudoscalar leptoproduction cross sections. 
For Reggeons, the $1^{--}$ does not couple at all to the longitudinal photon, while the axial vector $1^{+-}$ does through helicity flip~\cite{GolOwe}. Guided by these observations~\cite{Ahmad}, we assume the hard part depends on whether the exchange quantum numbers are in the vector or axial vector series, thereby introducing orbital angular momentum into the model. We use $Q^2$ dependent electromagnetic ``transition'' form factors for vector or axial vector quantum numbers going to a pion. We calculate these using PQCD for $q+\bar{q} + \gamma^*(Q^2) \rightarrow q+\bar{q}$ and a standard $z-$dependent pion wave function, convoluted in an impact parameter representation that allows orbital contributions to be easily implemented. 

With our model for the chiral odd, spin-dependent GPDs and these transition form factors, we can obtain the full range of cross sections and asymmetries in kinematic regimes that coincide with ongoing JLab experiments. We are able to predict the important transverse photon contributions to the observables~\cite{Ahmad}. Preliminary versions of this program have been presented and  further details will soon appear. A similar emphasis on chiral odd contributions has recently been proposed~\cite{GolKro}.

\section{Dispersion Relations and Partonic Interpretation of GPDs}

At the heart our understanding of the role of GPDs in exclusive leptoproduction reactions are the analyticity properties of the amplitudes. We have examined the applicability of Dispersion Relations (DRs) to the GPD formulation of DVCS. Unitarity and completeness are crucial ingredients in establishing analytic properties of the amplitudes. 
The amplitudes are analytic in energy variables, which allows the amplitudes (``Compton Form Factors'' or CFF's) to satisfy DRs relating real and imaginary parts. The imaginary part of a CFF is given by the GPD evaluated at the kinematic point where the returning quark has only transverse momentum relative to the nucleon direction. Then the DR can determine the real part thereby. However, at non-zero momentum transfer the DRs require integration over unphysical regions of the variables and that region is considerable - the real parts must still be measured by using interference with the Bethe-Heitler contribution~\cite{GolLiu}. 

We have also investigated the analyticity in the $X < \zeta$ region (the ERBL region), which conventionally is described as a quark-antiquark distribution in the proton. We extended the derivation of the parton model from  connected matrix elements for  non-local quark and gluon field operators in inclusive hard processes~\cite{Jaffe} to the non-forward GPDs~\cite{GolLiu2}. At leading twist, the kinematics require semi-disconnected amplitudes, {\it i.e.} vacuum fluctuations, that vitiate the partonic interpretation. In order to restore the sensible partonic picture it is necessary to include gluon exchange, appearing as an initial or final state interaction that ``dresses'' the struck or returning quark.

\section{Conclusion}
$\pi^0$ electroproduction provides a window into transversity. Chiral odd GPDs are essential because transverse $\gamma^*$ have a big role. A broad vista of spin phenomenology is opened up.

Finally some words about travel: I managed to fly on the only non-stop from Toronto to Rome - all other flights to Europe were cancelled. The next morning the Rome train station was jam-packed with thousands of people shoving in lines to the ticket counters. After waiting for an hour, I left the line, ran to the express train to Firenze without a ticket, got on against the advice of a conductor, and found an empty seat in first class. I had to pay a small fine, but arrived in Firenze in 1 1/2 hours. Amazing luck!
I am very grateful to the organizers for all the juggling of the schedule and the improvising to include many remote presentations. A great success under extraordinary circumstances!

\end{document}